\documentclass[apj]{emulateapj}

\slugcomment{To appear in ApJ}
\shortauthors{Kitayama \& Yoshida}
\shorttitle{Early SNRs}



\newcommand{\simgt}{\lower.5ex\hbox{$\; \buildrel > \over \sim \;$}}
\newcommand{\simlt}{\lower.5ex\hbox{$\; \buildrel < \over \sim \;$}}

\newcommand{\msun}{\,{\rm M}_{\odot}}
\newcommand{\HII}{H{\sc ii} }
\newcommand{\HI}{H{\sc i} }
\newcommand{\HeII}{He{\sc ii} }

\begin{document}

\title{Supernova explosions in the early universe: Evolution
of radiative remnants and the halo destruction efficiency}

\author{Tetsu Kitayama}
\affil{Department of Physics, Toho University, Funabashi,
Chiba 274-8510 Japan;  kitayama@ph.sci.toho-u.ac.jp}
\and
\author{Naoki Yoshida}
\affil{Department of Physics and Astrophysics, Nagoya University,
Nagoya 464-8064, Japan}

\begin{abstract}

We study the evolution of supernova (SN) remnants of the first stars,
taking proper account of the radiative feedback of the progenitor stars
on the surroundings.  We carry out a series of one-dimensional
hydrodynamic simulations with radiative cooling, starting from initial
configurations that are drawn from the results of our earlier radiation
hydrodynamic simulations of the first \HII regions.  We primarily
consider explosion energies $E_{\rm SN}=10^{51}-10^{53}$ erg,
appropriate for various types of a single Population III supernova, and
also explore cases with greater energy to model multiple explosions.  In
low-mass ($\simlt 10^6 \msun$) halos, the stellar radiation
significantly reduces the ambient gas density prior to the SN
explosion. The blastwave quickly propagates over the halo's virial
radius, leading to complete evacuation of the gas even with the input
energy of $10^{50}$ erg. We find that a large fraction of the remnant's
thermal energy is lost in $10^5-10^7$ yr by line cooling, whereas, for
larger explosion energies, the remnant expands even more rapidly with
decreasing interior density, and cools predominantly via inverse Compton
process, adding bulk of the energy to cosmic microwave background
photons.  In higher mass ($\sim 10^7 \msun)$ halos, the gas density near
the explosion site remains high ($\simgt 10^4$ cm$^{-3}$) and the SN
shock is heavily confined; the thermal energy of the remnant is quickly
radiated away by free-free emission, even if the total input energy
exceeds the binding energy of halos by two orders of magnitude. We show
that the efficiency of halo destruction is determined not only by the
explosion energy but also by the gas density profile, and thus
controlled by radiative feedback prior to the explosion.  We compute the
emissivity of the remnants in various energy ranges and examine the
detectability by future observations.  Several implications of our
results for the formation of first quasars and second-generation stars
in the universe are also discussed.

\end{abstract}

\keywords{cosmology:theory --- stars:Population III --- supernovae}

\section{Introduction}

The cosmic Dark Ages ended when the first stars lit up the universe.
This `dawn' of the universe may have been accompanied by rather violent
events; first supernova (SN) explosions.  Various feedback effects on
the intergalactic medium (IGM) are caused by SN explosions, and the
importance of early SNe can be easily appreciated by noting that only
light elements were produced during the nucleosynthesis phase in the
early universe.  Heavier elements such as carbon and oxygen must have
been expelled by means of SN explosions at some early epoch to account
for various observations of metal enrichment at high redshifts.

In the standard cold dark matter (CDM) model, the first objects are
formed in low-mass halos (Couchman \& Rees 1986; Tegmark et al. 1997;
Yoshida et al. 2003). Even a single SN explosion can destruct such a
`mini-halo' and thus can cause a strong negative feedback effect: if the
halo gas is completely blown away by a SN explosion, the subsequent
star-formation is likely to be suppressed temporarily in the same
region.  On the other hand, if the shock-heated gas remains trapped
within a deep gravitational potential, gas cooling and condensation may
occur even more efficiently thereafter; heavy elements processed in the
precursor star is dispersed in the vicinity of the explosion site, and
then the overall gas cooling efficiency can be locally enhanced. The
strengths of these effects and the halo destruction efficiency are of
considerable cosmological interest, because they may largely control the
global cosmic star formation rate at very high redshifts.

Recent theoretical studies on the formation of primordial stars
consistently suggest that the first stars are rather massive (Abel,
Bryan \& Norman. 2002; Bromm, Coppi \& Larson 2002; Omukai \& Palla
2003).  If the first stars are indeed as massive as $\sim 200
M_{\odot}$, they end their lives as energetic SNe via the
pair-instability mechanism (e.g., Barkat, Rakavy, \& Sack 1967; Bond,
Arnett, \& Carr 1984; Fryer, Woosley, \& Heger 2001; Heger \& Woosley
2002), releasing a total energy of up to $\sim 10^{53}$~erg.  Such
energetic explosions in the early universe are thought to be violently
destructive: they expel the ambient gas out of the gravitational
potential well of small-mass dark matter halos, causing an almost
complete evacuation (Bromm, Yoshida, \& Hernquist 2003; Wada \&
Venkatesan 2003).  Since the very massive stars process a substantial
fraction of their mass into heavy elements, early SN explosions may
provide an efficient mechanism to pollute the surrounding IGM (Schneider
et al. 2002; Yoshida, Bromm \& Hernquist 2004; Qian \& Wasserburg 2005).

The physics of astrophysical blastwaves has been extensively studied in
various contexts (e.g., Ikeuchi 1981; Bertschinger 1985; Vishniac,
Ostriker, Bertschinger 1985; see also a comprehensive review by Ostriker
\& McKee 1988).  On a cosmological background, Ikeuchi (1981) suggested
energetic explosions in the early universe as a large-scale
star-formation and galaxy formation mechanism.  Carr, Bond \& Arnett
(1984) and Wandel (1985) studied shell fragmentation in a similar
scenario involving Population III SN explosions in the early universe.
Dekel \& Silk (1986) studied the impact of star-bursts and associated
SN-driven winds on the formation of dwarf galaxies, whereas Tegmark,
Silk \& Evrard (1993) suggested a reionization model by SN-driven
blastwaves. Pre-galactic outflows driven by multiple SNe are suggested
also as an IGM enrichment mechanism at high redshifts (Madau, Ferrara \&
Rees 2001; Mori, Ferrara \& Madau 2002).  More recently, Bromm et
al. (2003) carried out three-dimensional (3-D) simulations of the first
SN explosions in a CDM universe and studied the various feedback effects
caused by them.  Wada \& Venkatesan (2003) performed grid-based Eulerian
hydrodynamic simulations of SN explosions in isolated small galaxies and
showed that the expelled gas falls back to the potential well after
about the system's free-fall time.  While these previous works
consistently showed the destructive aspect of early SN explosions, they
either simulated a few specific cases or employed somewhat simplified
initial conditions.  The density profile around the explosion sites can
be more complex and critically control the efficiency of cooling of SNRs
(e.g. Terlevich et al. 1992).  Therefore, it is important to carry out
hydrodynamic simulations starting from proper, realistic initial
conditions.

In our earlier work (Kitayama, Yoshida, Susa \& Umemura 2004; hereafter
Paper I), we studied the evolution of cosmological \HII regions around
the first stars, by solving self-consistently radiative transfer,
non-equilibrium chemistry, and hydrodynamics.  We showed that the final
gas density profile depends sensitively on the halo mass. In
`mini-halos' with mass $\simlt 10^6 M_{\odot}$, the ionization front
quickly expands to a radius of over 1 kpc and the halo gas is
effectively evacuated with the mean density of the halo decreasing to
$\simlt 1$ cm$^{-3}$ (see also Whalen et al. 2004). In larger mass
($\simgt 10^7 M_{\odot}$) halos, the \HII region is confined well inside
the virial radius and the central density remains high at $> 10^4$
cm$^{-3}$.  Since SN explosions are triggered at the death of the
central star, the evolved profiles of density, temperature, and velocity
of these simulations should serve as appropriate initial conditions for
the studies of the SN explosions and subsequent remnant evolution.

In the present paper, we study the evolution of SNRs in the
high-redshift universe, using 1-D numerical simulations that start from
the final outputs of our earlier simulations of the first \HII regions
mentioned above.  We primarily consider point explosions caused by
Population III SN with a total explosion energy of $10^{51}-10^{53}$
erg.  We calculate the destruction efficiency and the rate of radiative
loss from the SNRs to address their cosmological implications. Our study
differs from previous works on astrophysical explosions in the following
points; 1) our simulations include the gravitational force exerted by
the host dark matter halo, and 2) we adopt realistic initial conditions
for Population III SN sites that are calculated by self-consistent
radiative transfer calculations. The simulations incorporate all the
relevant cooling processes at $T > 10^4$ K for a primordial gas,
including the inverse-Compton cooling which is important at $z>10$.  We
also examine several cases with an enormous explosion energy, $E_{\rm
tot} = 10^{54}$ erg, on the assumption that multiple massive stars form
in a single star-forming region and that they explode almost
simultaneously.  This situation may indeed be plausible in large mass
systems in which gas can cool efficiently via hydrogen atomic cooling.
Because the lifetime of massive stars (supernova progenitors) is
typically comparable to or shorter than the evolution time scale of
SNRs, we approximate such simultaneous multiple explosions by a single
explosion with a large total energy.
 
Throughout the paper, we work with a $\Lambda$-dominated cosmology with
matter density $\Omega_{\rm M}=0.3$, cosmological constant
$\Omega_{\Lambda}=0.7$, the Hubble constant $h=0.7$ in units of $100$ km
s$^{-1}$Mpc$^{-1}$, and baryon density $\Omega_{\rm b}=0.05$.

\section{The simulations}
\subsection{The code}

We use the hydrodynamics code of Kitayama \& Ikeuchi (2000) and Kitayama
et al. (2000, 2001, 2004).  It employs the second-order Lagrangian
finite-difference scheme in spherically symmetric geometry (Bowers \&
Wilson 1991), and treats self-consistently gravitational force,
hydrodynamics, and radiative cooling. The basic equations are presented
in Sec. 2 of Kitayama \& Ikeuchi (2000). We adopt an artificial
viscosity formulation of Caramana, Shashkov \& Whalen (1998), designed
to distinguish between shock-wave and uniform compression using an
advection limiter.

The gas is assumed to have the primordial composition with hydrogen and
helium mass fractions, $X=0.76$ and $Y=0.24$, respectively. The atomic
cooling rates are taken from the compilation of Fukugita \& Kawasaki
(1994).  Since our primary interests lie in the evolution of SNRs at
temperatures $T > 10^4$ K until the formation of cooled dense shells,
formation of molecules and cooling by them at lower temperatures are not
included in the present paper to avoid computational difficulties.  As
discussed in Appendix, we can assume that electron-ion equilibration is
a reasonable assumption in the present calculation.  The results of
basic code tests relevant to SNR evolution are presented in
Appendix. From the results, we conclude that our Lagrangian code
accurately reproduces analytic solutions and that it also captures shock
regions remarkably well.

\subsection{Initial conditions}

Supernova explosions are triggered {\it after} the progenitor stars have
shined for their lifetime. Consequently the ambient gas density and
temperature profiles are significantly modified by their radiation.  As
shown in Paper I, the resulting density and temperature profiles are
indeed quite complex, being dependent on the central source luminosity,
density profile, and the host halo mass.  We thus take the final output
of our previous simulations for the ionization front (I-front)
propagation as initial configurations for the present explosion
simulations.

\begin{figure}
\hspace*{-2.7cm} \plotone{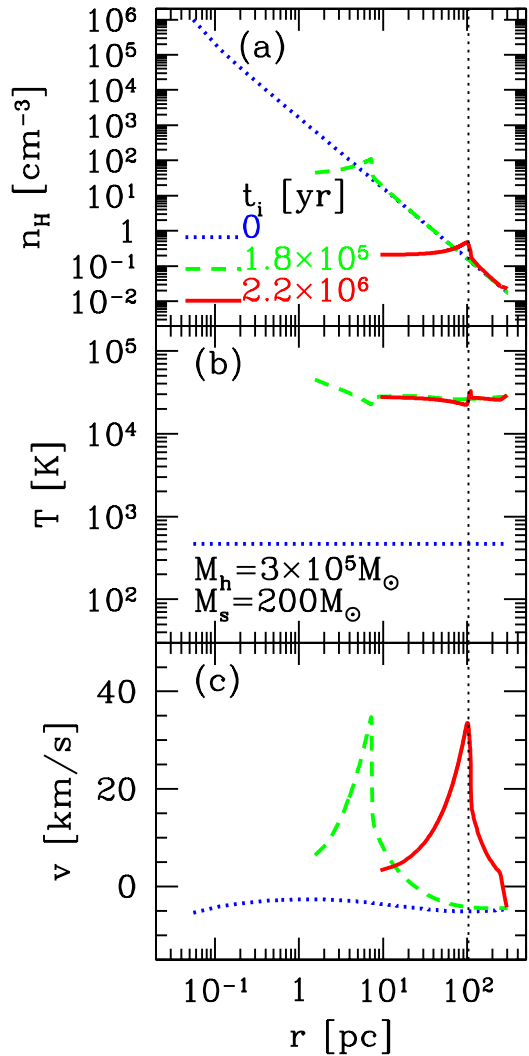} 
\caption{Expansion of the \HII region around a $M_{\rm
s}=200$ M$_\odot$ star embedded in a halo with $M_{\rm h}= 3.2\times
10^5$ M$_\odot$ and $z_{\rm c}=20$; (a) hydrogen density, (b) gas
temperature, and (c) outward velocity at $t_i=0$ (dotted lines),
$1.8\times 10^5$ (dashed) and $2.2 \times 10^6$ yr (solid), where $t_i$
denotes the time elapsed since the birth of the massive star. The
vertical thin dotted line indicates the halo virial radius.
\label{fig:prof0}}
\end{figure}

Unless otherwise stated, we adopt the fiducial set up in Paper I; a
single progenitor star of mass $M_{\rm s} = 200 M_{\odot}$ is placed at
the center of a halo collapsed at redshift $z_{\rm c}=20$.  The gas
density profile prior to the birth of the star is $\rho \propto r^{-2}$
and is subsequently altered by the stellar radiation during its lifetime
2.2 Myr (Schaerer 2002).  Figure \ref{fig:prof0} shows a characteristic
example of such runs in the case of halo mass $M_{\rm h}=3.2 \times 10^5
\msun$. The whole halo is ionized promptly and the gas inside the
central $\sim 100$ pc is evacuated within the lifetime of the massive
star, with the inner gas density being $\sim 0.1-1 {\rm cm}^3$.  For
larger mass halos with $M_{\rm h} \simgt 10^7 \msun$, on the other hand,
the I-front expansion is greatly hindered. Gas infall from the outer
envelope continues in such cases and the density within the compact \HII
region remains high at $>10^4$ cm$^{-3}$ (Fig. 2 of Paper I).

\epsscale{2.0}

When making the initial conditions from the outputs of the \HII region
calculations, we need to add extra shells and then interpolate the
physical quantities in order to resolve thin cooling shocks.  We adopt
the following second-order re-zoning scheme so that their monotonicity
is maintained as well as mass, momentum and thermal energy are conserved
in each shell.  First, gas density is expressed as a function of enclosed
volume as $\rho(V)$. For the $i$-th shell, we denote its center of
volume by $V_i$ and assign the mean gas density of the shell  
$\bar{\rho}_i$ to $V_i$, i.e. $\rho(V_i)=\bar{\rho}_i$. Second, the
density gradient within the $i$-th shell is defined by
\begin{equation}
(d\rho/dV)_i = \mbox{minmod} (s_{i-1/2},s_{i+1/2}), 
\end{equation}
where 
\begin{eqnarray}
s_{i-1/2} &=& [\rho(V_i)-\rho(V_{i-1})]/(V_i - V_{i-1}), \\
\mbox{minmod}(x,y) &=& 
\left\{\begin{array}{ll}
   \mbox{sgn}(x) \min(|x|,|y|) & \mbox{if $xy >0$}, \\
    0  & \mbox{otherwise.} \\
    \end{array} \right.  
\end{eqnarray}
The gas density at an arbitrary point within the $i$-th shell is
computed by $\rho(V) = \rho(V_i) + (d\rho/dV)_i \times
(V-V_i)$. Finally, similar procedures are repeated for the internal
energy per unit mass and for the velocity, except that they are expressed as
a function of enclosed mass as $u(m)$ and $v(m)$, respectively. 

For comparison, we also perform some runs without the radiative feedback
prior to the SN explosion. Specifically, we adopt the same initial
configurations as those in Paper I; the density profile follows a pure
power-law with $\rho \propto r^{-2}$, the gas is isothermal and
infalling with the free-fall velocity (dotted lines shown as initial
configuration in Fig. \ref{fig:prof0}).  We adopt this somewhat
artificial configuration in order to examine how the final results of
our explosion simulations are affected by the initial density profiles.

All the simulations are initiated when the SN ejecta, assumed to have
$M_{\rm ej}= M_{\rm s}$, has swept up the surrounding mass equal to its
own, i.e., at the end of the initial free-expansion phase (see
Appendix). This phase ends roughly at $t_{\rm sw}$ given in equation
($\ref{eqsw}$) after the explosion. The extent of the inner-most shell 
is so chosen that the enclosed mass is $M_{\rm ej}$.  We then add the
ejecta mass $M_{\rm ej}$ and the thermal energy $E_{\rm SN}$ to this
cell assuming that the ejecta and the swept-up material are fully
thermalized. In some cases, the density of the inner-most shell  is
larger than $\sim 10^5 {\rm cm}^{-3}$, its radiative cooling time gets
smaller than $t_{\rm sw}$, and the above assumption may not be valid
(Terlevich et al. 1992; see also Appendix). We therefore adopt the
following prescription for the initial energy input. To the crudest
approximation, the energy lost before the Sedov phase is $E_{\rm
rad}(t_{\rm sw}) = \dot{E}(t_{\rm sw}) \times t_{\rm sw}$, where
$\dot{E}(t_{\rm sw})$ is the rate of energy loss in the inner-most shell 
computed under the above assumption. If $E_{\rm SN} - E_{\rm rad}(t_{\rm
sw})$ is still greater than zero, we add it to the inner-most shell 
instead of $E_{\rm SN}$ and start the simulation. If it is less than
zero, on the other hand, we regard the SN shock as being stalled during
the initial phase and do not perform the simulation.  More detailed
discussion on the relevant time-scales is presented in Appendix.

We vary the input explosion energy $E_{\rm SN}$ over the range
$10^{50}-10^{53}$ erg to account for various possibilities on the fate
of massive stars; Type II SN, pair-instability SN, or by
hypernovae (Heger \& Woosley 2002; Umeda \& Nomoto 2002).  We also
explore the cases where 10 progenitor stars with $M_{\rm s} = 200
M_{\odot}$ have produced an \HII region and explode with a total energy
of $E_{\rm SN}=10^{54}$ erg.  Since the lifetime of the massive
stars, a few million years, is shorter than the characteristic evolution
time of the SNR, we approximate the multiple SNe as a single explosion at
the center with $E_{\rm SN}=10^{54}$ erg.

The outer boundary is taken at $1-100$ times the virial radius depending
on the evolution. Since we do not know {\it a priori} the extent to
which the blastwave can propagate, this procedure required us to carry
out some of the simulations more than once. We set progressively larger
radius for the outer boundary and made sure that it encloses the SN
shock after $10^7$ yr and also that the boundary radius does not affect
the result within the shock radius.  Whenever necessary, we extrapolate
the gas density to the envelope assuming $\rho \propto r^{-2}$.  We
assume that the internal energy and velocity are constant in the
extrapolated region in accord with the adopted isothermal density
profile.  The pressure outside the outer boundary is set by
extrapolating the pressure inside the boundary by a quadratic
polynomial.

The gas shells are spaced such that the shell mass changes by a constant
ratio, less than 1\%, between the adjacent shells. The total number of
shells is $N=1000$. We have checked that our results remain unchanged
when the shell number is doubled or halved.

For the dark matter component, we assume that the density profile is
given by a Navarro, Frenk \& White (1997) profile as in Paper I.  We
follow Bullock et al. (2001) to determine the concentration parameter of
a halo with total mass $M_{\rm h}$ collapsing at redshift $z_{\rm
c}$, by extrapolating their formula to the lower halo masses and the
higher redshifts. The dark matter density is normalized so that the dark
matter mass within the virial radius is equal to $M_{\rm DM}= (1 -
\Omega_{\rm B}/\Omega_{\rm M}) M_{\rm h}$. As we focus on the
evolution within $10^7$yr after the SN explosion, less than the halo
dynamical time at $z=20$, we assume that the dark matter density is
unchanged throughout the simulation.

\section{Results}

\subsection{Evolution of early SNRs} 
\label{sec:ev}

\begin{figure}
\hspace*{-2.7cm} 
\plotone{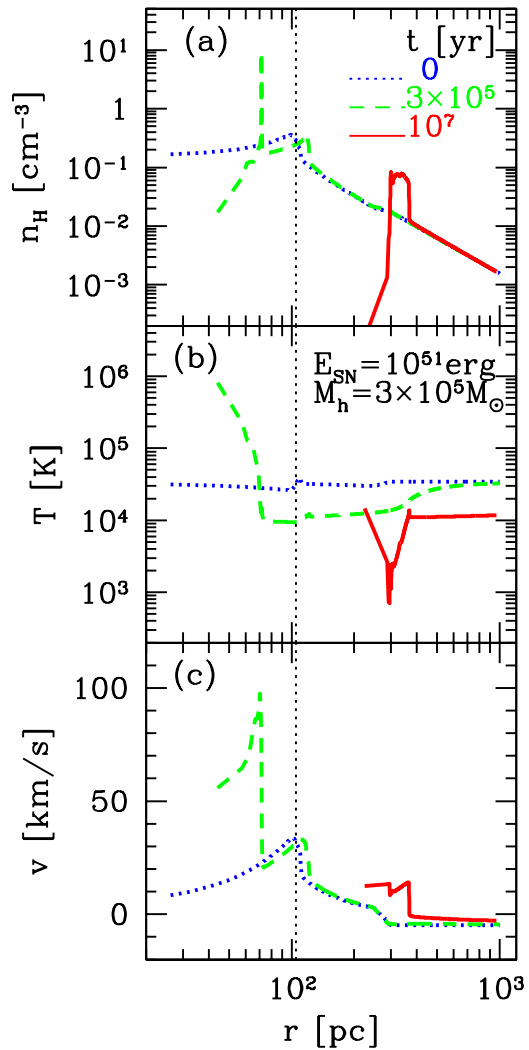} \caption{Evolution of an early SNR in the case of
$E_{\rm SN}=10^{51}$ erg, $M_{\rm h}= 3.2 \times 10^5$ M$_\odot$, and
$M_{\rm s}=200$ M$_\odot$; (a) hydrogen density, (b) gas temperature,
and (c) outward velocity at $t=0$ (dotted lines), $3\times 10^5$
(dashed) and $10^7$ yr (solid), where $t$ denotes the time elapsed since
the end of the free expansion stage.  The vertical thin dotted line
indicates the halo virial radius.  \label{fig:prof1} }
\end{figure}

\begin{figure}
\hspace*{-3.3cm} 
\plotone{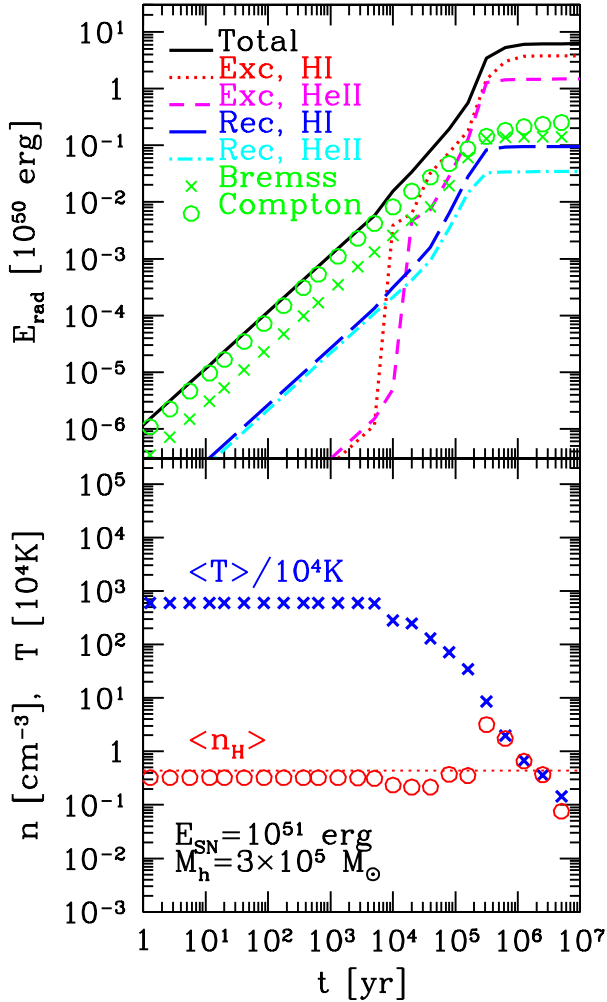} \caption{(a) Cumulative energy radiated from the SN
shocked region for the run presented in Figure \ref{fig:prof1}.  Lines
and symbols indicate the contributions to the total radiation energy
(solid lines) from excitation of \HI (dotted) and \HeII (short dashed),
recombination to \HI (long dashed) and \HeII (dot dashed), thermal
bremsstrahlung (crosses) and Compton scattering with CMB photons
(circles).  (b) Mass-weighted average temperature $\left< T\right>$
(crosses) and average density $\left< n_{\rm H}\right>$ (circles) of the
SN shocked region. The horizontal dotted line indicates the initial
(before radiative feedback) mean density of the gas within the virial radius.
\label{fig:rad1} }
\end{figure}

As a characteristic example of an early SNR evolution, we plot in Figure
\ref{fig:prof1} the radial profiles for the run with $E_{\rm
SN}=10^{51}$erg, $M_{\rm h}= 3.2 \times 10^5 M_{\odot}$ and $z_{\rm
c}=20$.  In what follows, we denote by $t$ the time elapsed since the
end of the initial free-expansion stage. Prior to the SN explosion, the
central massive star has ionized the gas beyond the virial radius and
the associated shock has swept away the surrounding gas. The central gas
density is $\sim 0.2$ cm$^{-3}$ (Fig.\ref{fig:prof0}) at the time when
the explosion is triggered.  This greatly helps the propagation of the
newly formed SN shock front.  It quickly catches up with the foregoing
shock generated during the \HII region formation (see the double peaks
in density at $t=3\times 10^5$ yr).  The halo gas is evacuated almost
completely by $t=10^7$ yr, and the swept-up material resides in a dense
`shell' at $r\sim 300$ pc.  We found that the evolution of the SN
blastwave before it merges with the foregoing shock is well described by
the Sedov self-similar solution for a constant ambient density.

Figure \ref{fig:rad1} shows the cumulative energy radiated from inside
the SN shock front and the {\it mass-weighted} mean temperature and
density of the SN shocked region for the same run.  At $t < 10^4$yr, the
mean temperature of the shocked region is $> 10^6$K and the dominant
radiative processes are free-free and Compton cooling. The total
emission {\it rate} is the largest at $t \sim 10^5$yr, when
the dense shell forms (eq. \ref{eq:tsg}) and the mean density of the
shocked region is the largest.  Bulk of the thermal energy is then lost
via collisional excitation cooling of He$^{+}$ and H. Thereafter the SNR
enters the pressure-driven expansion stage.

We note here that our code does not solve the radiative transfer of
photons emitted from the remnant itself.  As the blast wave is induced
by the pressure gap much greater than that attained by photoionization,
the overall effect of the radiative precursor on the dynamics of
blastwaves is expected to be small. To check this, we have performed a
test run, constraining the temperature of the surrounding gas to be
above $10^4$ K to approximate the pre-ionization ahead of the SN shock.
For the same set of parameters as in Figure \ref{fig:prof1}, the shock
radius at $t=10^7$yr changes by only $\sim 30\%$.  The difference will
be even smaller for stronger blastwaves with larger $E_{\rm SN}$.  Thus
we neglect the effect of radiative precursors in the rest of the
paper. We refer the reader to a detailed study of the radiative transfer
and the emission spectra by Shull \& McKee (1979).

\begin{figure}
\hspace*{-2.7cm} \plotone{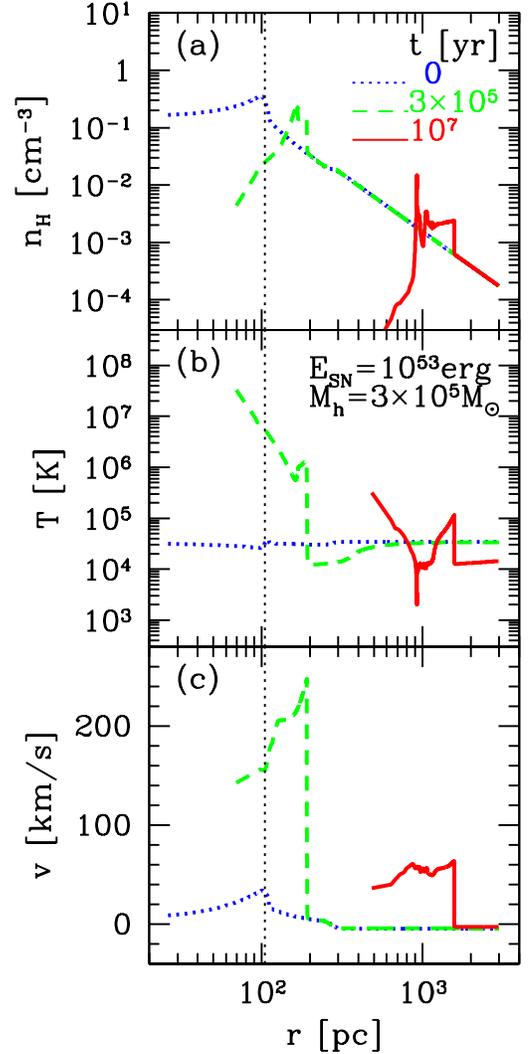} 
\caption{Same as Figure \ref{fig:prof1} except for 
$E_{\rm SN}=10^{53}$ erg.
\label{fig:prof2} }
\end{figure}

\begin{figure}
\hspace*{-3.3cm} 
\plotone{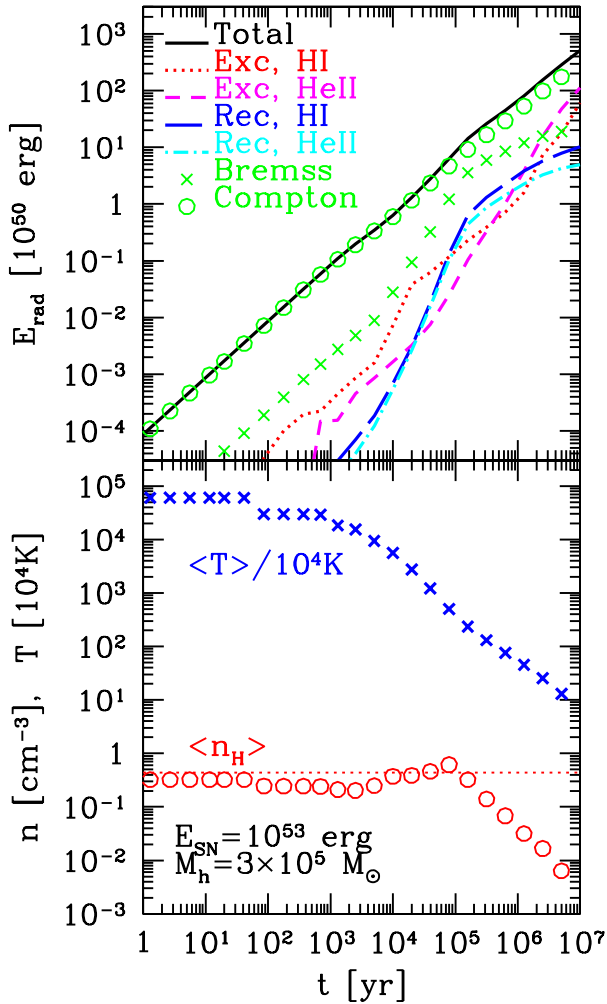} 
\caption{Same as \ref{fig:rad1} except for 
$E_{\rm SN}=10^{53}$ erg.
\label{fig:rad2} }
\end{figure}

The case of a larger total SN energy $E_{\rm SN}=10^{53}$ erg is
displayed in Figures \ref{fig:prof2} and \ref{fig:rad2}. The gas is
quickly evacuated and the shock front exceeds the virial radius at $t
\sim 10^5$yr. The remnant temperature decreases initially via adiabatic
expansion rather than radiative cooling.  Due both to high temperature
and low density of the shocked region, Compton cooling ($\propto n_e
T_e$) dominates over the other radiative processes throughout the
evolution.  In this case, approximately 30\% of the explosion energy is
lost via the Compton cooling.  Clearly, explosion energy can be
efficiently transferred to cosmic microwave background (CMB) photons
only if the explosion energy is large and/or the gas density is small.

The situation drastically changes if there were no I-front expansion
prior to the SN explosion. If the initial density profile of the run
shown in Figure \ref{fig:prof1} is a pure power-law with $\rho \propto
r^{-2}$, the cooling time of the inner-most shell gets smaller than
$t_{\rm sw}$ and the ejected energy is rapidly lost during the
free-expansion stage, i.e. $E_{\rm rad}(t_{\rm sw}) > E_{\rm SN}$.
Accordingly, the blastwave {\it stalls} in the dense environment and the
halo gas in the outer envelope will not be disturbed.  This is the case
even if the ejected SN energy is much greater than the binding energy,
$E_{\rm bin} \sim 10^{49}$erg, for baryons within the virial radius of a
$M_{\rm h}= 3.2\times 10^5 \msun$ halo. It clearly shows the importance
of setting-up appropriate initial configurations in quantifying the
degree of SN feedback. We emphasize that simple analytic estimates based
on explosion energy to binding energy ratio are inappropriate.

\begin{figure}
\hspace*{-2.7cm} \plotone{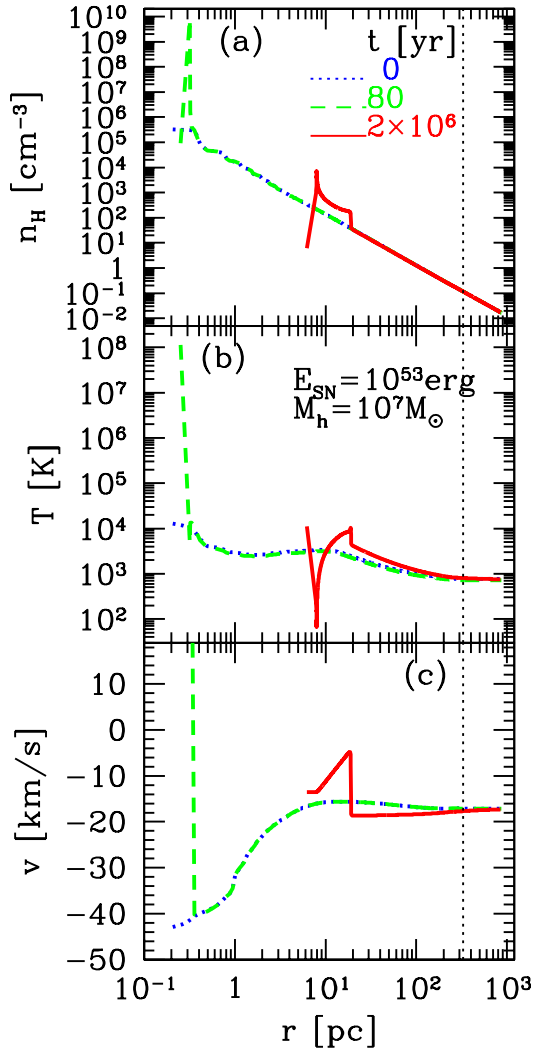} 
\caption{Same as Figure \ref{fig:prof1} except for 
$E_{\rm SN}=10^{53}$erg and $M_{\rm h}=10^7\msun$.  
Note that we plot profiles
for a broader range of radial distance than in Figure \ref{fig:prof1}
and \ref{fig:prof2}. 
The peak velocity at $t=80$ yr is 1400 km s$^{-1}$.
\label{fig:prof3} }
\end{figure}

\begin{figure}
\hspace*{-3.3cm} \plotone{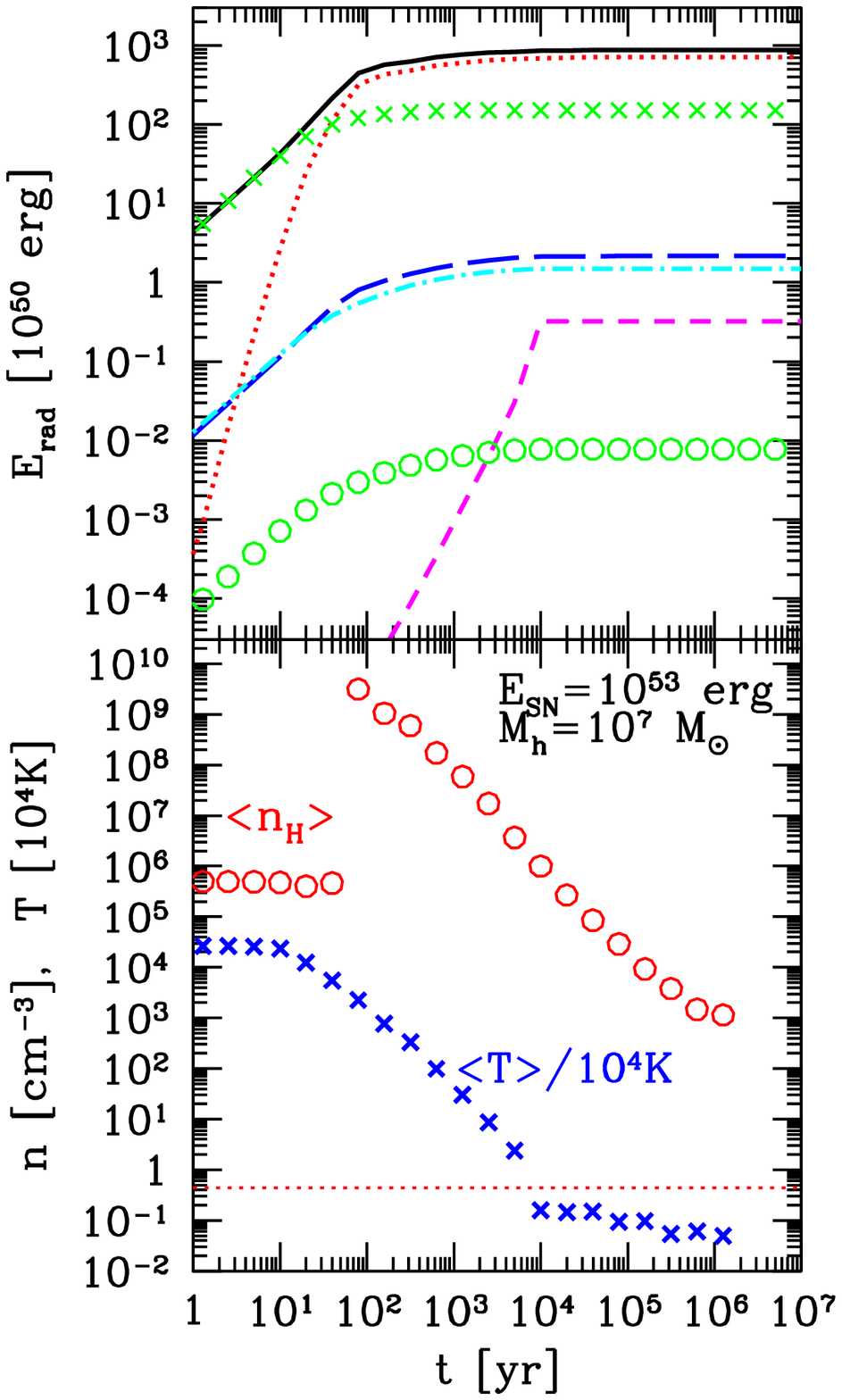} 
\caption{Same as \ref{fig:rad1} except for 
$E_{\rm SN}=10^{53}$ erg and $M_{\rm h}=10^7\msun$.
\label{fig:rad3} }
\end{figure}

Corresponding cases of no I-front expansion are realized in a halo as
large as $M_{\rm h} = 10^7\msun$. As shown in Figure \ref{fig:prof3},
the central gas density remains at $>10^5$ cm$^{-3}$ even when the
radiation from the massive star has photo-ionized and photo-heated the
gas (see Paper I).  The ejected SN energy is lost mainly via free-free
emission from the shock interior and collisional excitation cooling from
just behind the shock front (Fig. \ref{fig:rad3}). The latter process
efficiently converts kinetic energy into radiation in the present case;
the gas behind the blastwave is shock-heated but radiates quickly at $T
\sim 10^4$ K.  This process dominates the total cooling rate at $t>30$
yr.  Due to high expansion velocity and high cooling efficiency, an
extremely dense shell forms at $t=80$ yr.  The shock expands only to
$\sim 10$ pc in $t=2 \times 10^6$ yr and stalls thereafter, being well
inside the virial radius.  Gravitational force by dark matter pulls the
gas to the center, and the mass inflow is eventually recovered.  The
mass deposition rate inside $r=1$ pc is $\sim 0.1\msun$ yr$^{-1}$ at
$t=10^7$ yr.  In order to evacuate the gas out of the virial radius,
multiple SN explosions with $E_{\rm SN}\sim 10^{54}$ erg is
required. Again this is much greater than the binding energy for baryons
within the virial radius, $E_{\rm bin} \sim 3 \times 10^{51}$ erg.

It is illustrative to compare our 1-D calculation results with those of
3-D simulations. We have attempted a comparison with the Smoothed
Particle Hydrodynamis simulations of Bromm et al (2003) as
follows. First, we set $E_{\rm SN}=10^{51}$ and $10^{53}$ erg in a
$M_{\rm h}=10^6\msun$ halo collapsing at $z=20$ as found in the parent
cosmological simulation of Bromm et al (2003).  Since they used an
approximate treatment for the radiative feedback prior to the explosion,
we decided to use the resulting gas density profile, rather than
directly modeling the radiative feedback in their manner.  The gas
density profile is given such that there is a flat core with $n_{\rm
H}=1$ cm$^{-3}$ within the radius $r_{\rm core}=20$ pc and the envelope
$n_{\rm H} \propto r^{-2}$ beyond that, as described in Bromm et al
(2003). Since all the gas in the halo is ionized within the lifetime of
the massive star, we assume that the initial temperature is uniform at
$T=3\times 10^4$ K.  To allow sufficient spatial resolution near the
center, the mass of the inner-most shell is set to be $1 \msun$,
corresponding to the initial radius of 2.5 pc for this cell. The shell
mass increases by $1\%$ between the adjacent shells toward the envelope.
We ran the simulations starting from this configuration. The
radial positions of the dense shell are found to be $r_{\rm sh}=13$, 33,
89, and 214 pc at $t=10^4$, $10^5$, $10^6$, and $10^7$ yr for $E_{\rm
SN}=10^{51}$ erg, and $r_{\rm sh}= 13$, 34, 130 and 570 pc at $t=10^3$,
$10^4$, $10^5$, and $10^6$ yr for $E_{\rm SN}=10^{53}$ erg,
respectively. Overall the evolution of the shock radius agrees well with
the results of Bromm et al (2003). At $r_{\rm sh} \simlt 30$ pc, our
calculations predict about a few times larger shock radius (i.e. faster
expansion) than that plotted in Fig.~1 of Bromm et al (2003).  We have
checked that our simulations accurately reproduce the analytical
solutions for the blast wave propagation in the Sedov phase (Appendix).
Thus the disagreement at the small radii could be due to the limited
spatial resolution of the 3-D simulations, and/or due to details in the
way the initial explosion energy is deposited.  Given that the density
and velocity fields are likely more complex in the 3-D simulations,
performing more accurate comparison would be difficult, and we do not
attempt to do so further.  We emphasize that it is indeed remarkable
that the late-time evolution of the SNR is quite similar in both cases.

\subsection{Radiative loss from early SNR}
\label{sec:rad}

\begin{figure}
\plotone{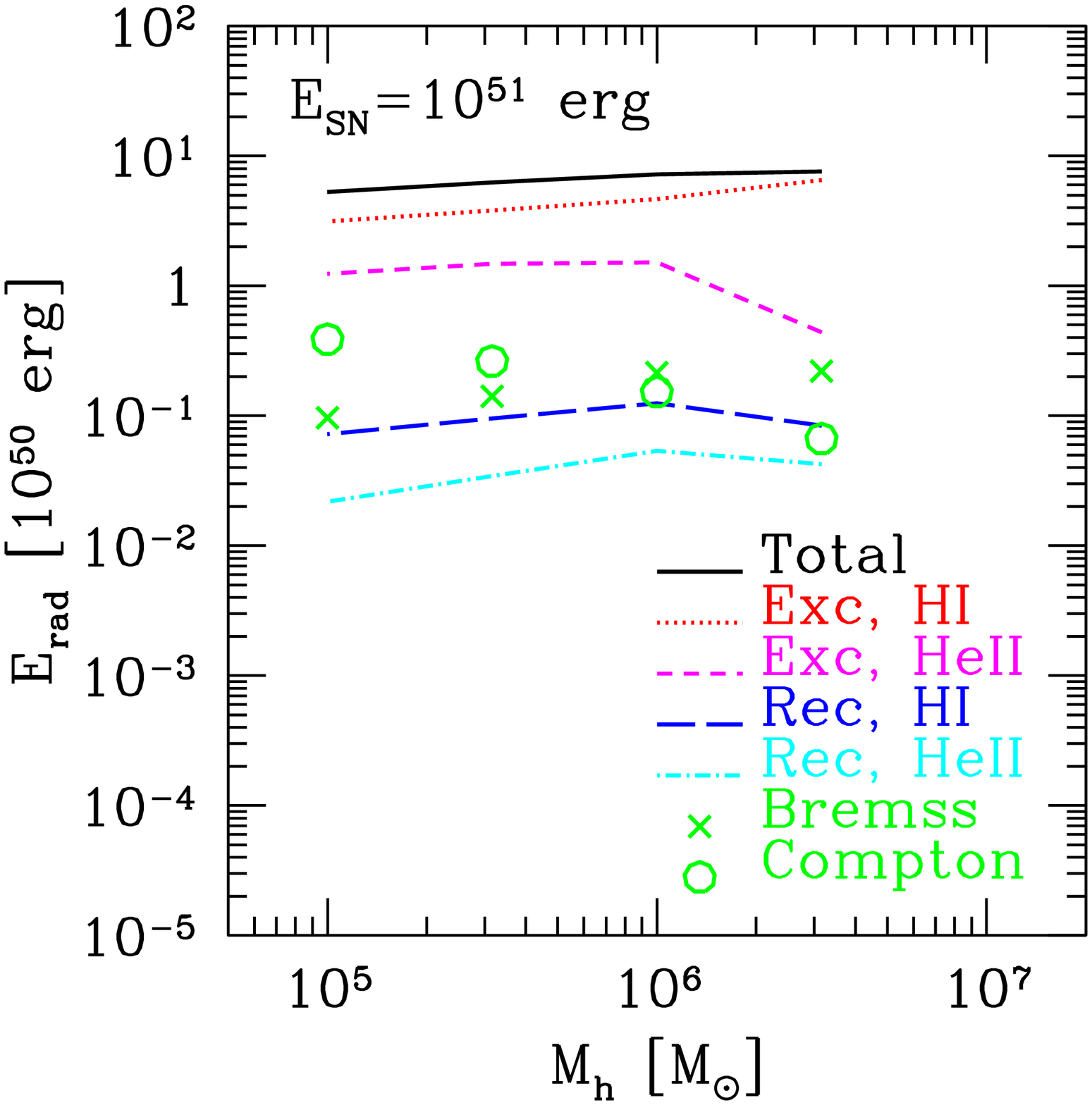}\plotone{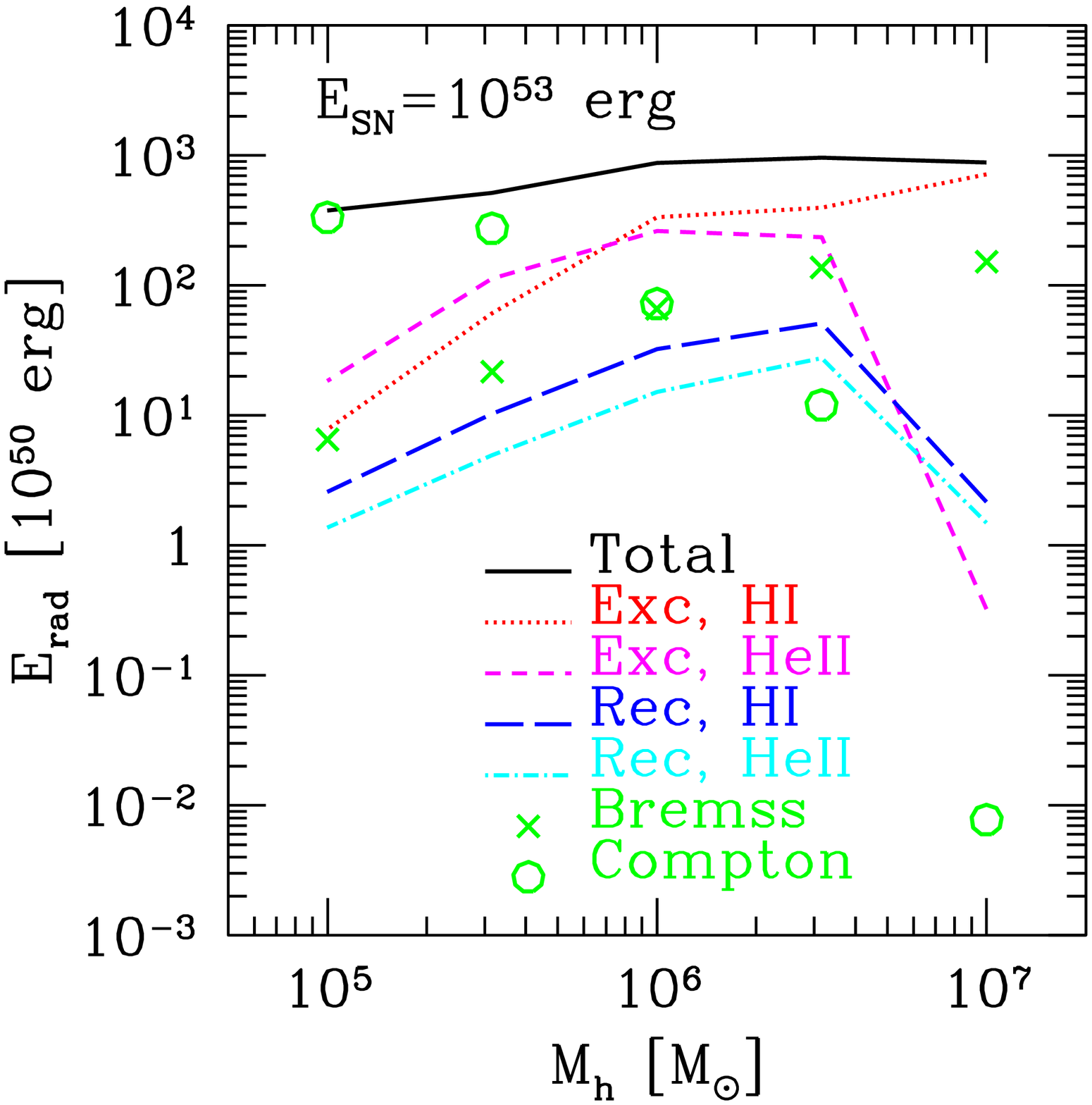} \\\plotone{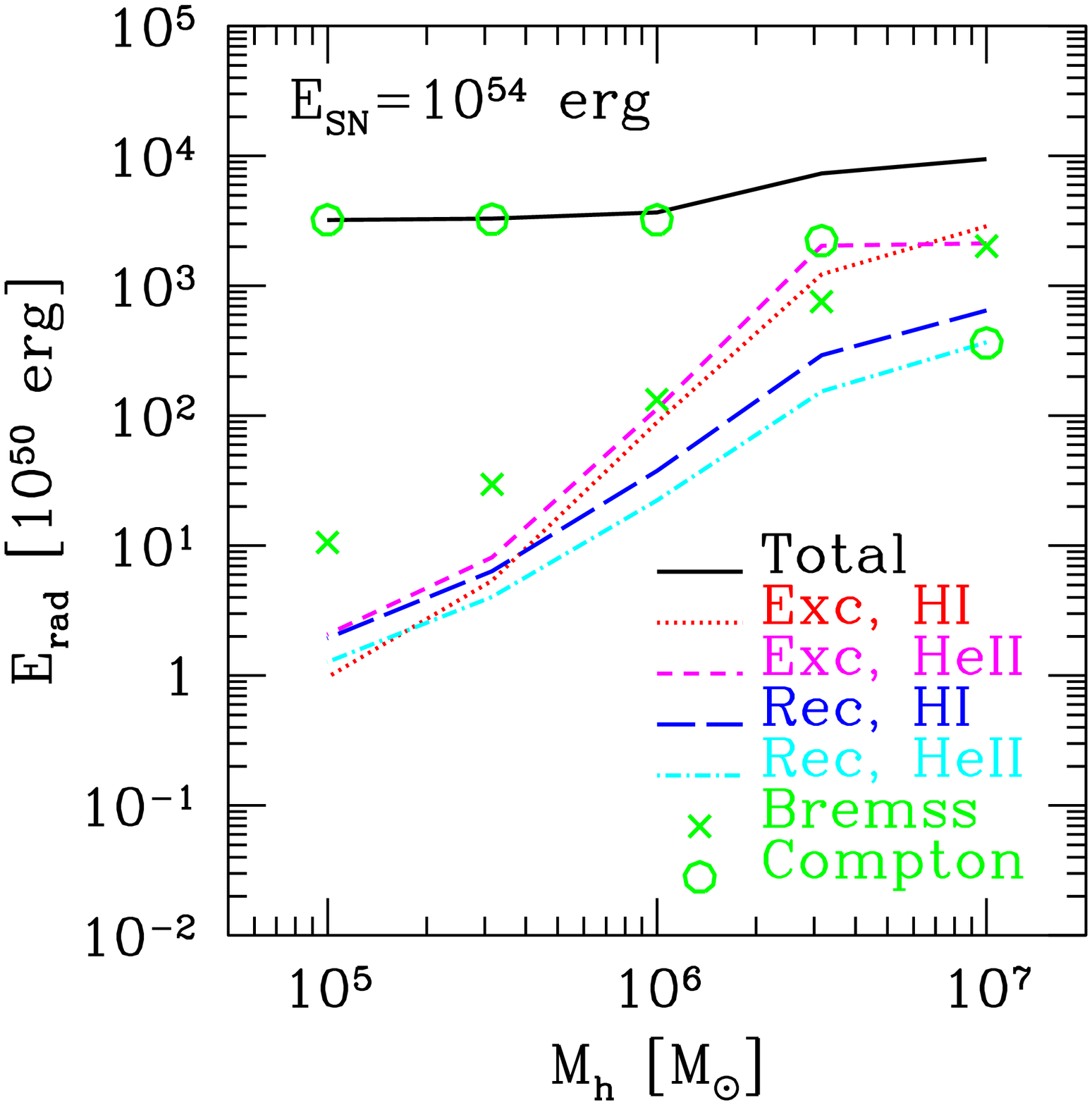} \caption{The
energy radiated from the SN shocked region within $10^7$ yr for $E_{\rm
SN}=10^{51}$ erg (top panel), $E_{\rm SN}=10^{53}$ erg (middle 
panel), and $E_{\rm SN}=10^{54}$ erg (bottom panel), as a function of
halo mass. \label{fig:radm} }
\end{figure}

We have shown that the dominant cooling process during the SNR evolution
varies with the ambient gas density and the explosion energy.  In Figure
\ref{fig:radm}, we plot the radiated energy from the SN shocked region
within $10^7$ yr as a function of halo mass. The three panels differ in
the explosion energy, as indicated in the figure.  Note that, for
low-mass halos, the total radiated energy is less than $E_{\rm SN}$
because a part of the SN energy is still in the form of kinetic energy
to drive the shell expansion. We find that there are at least three key
radiative processes in the evolution of high-redshift SNRs.

\epsscale{1.0}

\begin{figure}
\plotone{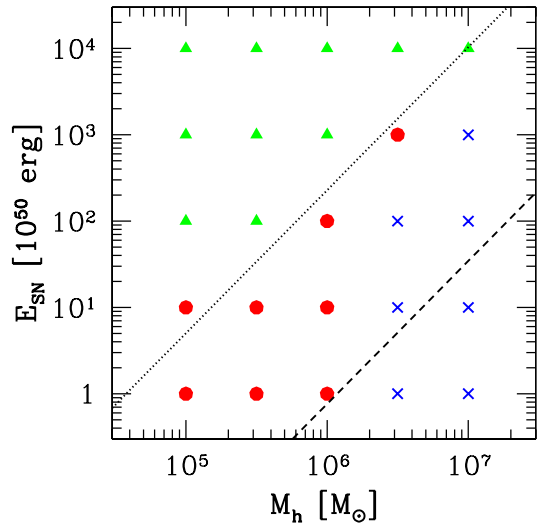} \caption{$E_{\rm SN}- M_{\rm h}$ diagram summarizing
the fate of host halos. Halos blown away even in the absence of initial
I-front expansion are marked by triangles, those blown away only in the
presence of I-front initial expansion by circles, and those not blown
away by crosses.  Dashed and dotted lines show the binding energy of the
gas for a given $M_{\rm h}$ and 300 times the same quantity,
respectively.  \label{fig:em} }
\end{figure}

First, collisional excitation cooling of hydrogen and helium is dominant
for high $M_{\rm h}$ or low $E_{\rm SN}$ (e.g., Figs \ref{fig:rad1} and
\ref{fig:rad3}).  It is particularly important in the later stage of the
SNR evolution, when the temperature drops from $10^7$ K to $10^5$ K and
a thin-shell appears.  The most prominent process is hydrogen Ly$\alpha$
emission. For $E_{\rm SN}=10^{51}$erg and $M_{\rm halo}= 3.2 \times 
10^5 \msun$, about $25 \%$ of the 
ejected energy is ultimately lost by the Ly$\alpha$
line in $5 \times 10^5$ yr, yielding an time-averaged luminosity
$L_{\alpha} \sim 10^{37}$erg s$^{-1}$.

Second, Compton cooling is dominant for either very low $M_{\rm h}$ or
very high $E_{\rm SN}$; in this case the remnant expands very rapidly
and cools mainly after its interior density has decreased significantly
(e.g., Fig.  \ref{fig:rad2}).  It is worth noting that these SNRs can
produce small angular scale fluctuations of the CMB via the
Sunyaev-Zel'dovich effect (e.g. Oh, Cooray \& Kamionkowski 2003).  In an
extreme case of $E_{\rm SN}=10^{54}$erg and $M_{\rm h} = 3.2 \times 10^5
\msun$, $\sim 30 \%$ of the ejected energy is lost by Compton cooling in
$10^7$ yr, yielding $\dot{E}_{\rm C} \sim 10^{39}$erg s$^{-1}$. The halo
virial radius corresponds to $0.06$ arcsec, and the mean $y$-parameter
within it is $3 \times 10^{-6}$ in the first $10^4$ yr and drops
thereafter as the gas temperature decreases. We stress that this is a
rather extreme case with the explosion of ten very massive stars within
a small mass halo.  The total energy lost via Compton cooling declines
rapidly for larger $M_{\rm h}$ or smaller $E_{\rm SN}$.  Thus the above
numbers should serve as the most optimistic estimates for the
Sunyaev-Zel'dovich effect caused by an individual SNRs.

Finally, free-free emission is important for massive halos
($>10^7\msun$) in which I-front expansion is suppressed.  A large
fraction of the explosion energy is lost immediately in the high density
and high temperature region near the center, as in the case shown in
Figs \ref{fig:prof3} and \ref{fig:rad3}. Although the mechanical
feedback is rather weak in this case, these SNRs can be significant
compact X-ray sources. In the case of $E_{\rm SN}=10^{53}$erg and
$M_{\rm h}=10^7 \msun$, for example, $\sim 15 \%$ of the ejected energy
is lost by free-free emission in 30 yr, yielding the rest-frame
luminosity of $L_{\rm X} \sim 10^{43}$erg s$^{-1}$ (Fig.
\ref{fig:rad3}).  The corresponding number for $E_{\rm SN}=10^{51}$erg
and $M_{\rm h}=3.2 \times 10^5 \msun$ is $\sim 1 \%$ in $3 \times 10^5$
yr (Fig. \ref{fig:rad1}), yielding $L_{\rm X} \sim 10^{36}$erg s$^{-1}$.  

\subsection{Criteria for the blow-away of halo gas}

It is our primary goal in the present paper to quantify the halo
destruction efficiency by SNe.  The fate of host halos are summarized in
Figure \ref{fig:em} for a wide range of halo mass and explosion energy.
We regard a halo to be `blown away', if more than 90\% of the gas
inside the virial radius is expelled within $10^7$ yr.  We have checked
that the result is insensitive to the specific choice of this threshold
between 50\% and 99\%.

Figure \ref{fig:em} shows a clear trend that greater SN energy is
required to blow away larger mass halos. It should be noted, however,
that the required SN energy is roughly 300 times larger than the total
binding energy for baryons in a given halo, if I-front expansion (and
hence the decrease of the gas density) prior to the explosion is
ineffective.  As discussed earlier, this is due to efficient cooling in
the high density ($>10^4$ cm$^{-3}$) environment near the halo
center. Once the I-front expansion is properly taken into account,
low-mass halos with $M_{\rm h} \simlt 10^6 \msun$ are blown away even by
a single SN with $E_{\rm SN}=10^{50}$ erg.  For higher mass halos
($M_{\rm h} \simgt 10^7 \msun$), the impact of I-front expansion weakens
sharply as shown in Paper I.

We summarize our findings as follows, emphasizing a close link between
the fate of halo gas classified in Figure \ref{fig:em} and the main
radiative processes described in Sec \ref{sec:rad}.  Compton cooling
tends to dominate over other processes in halos most readily blown away
(triangles).  These SNRs will cause distortion in the CMB spectrum via
the Sunyaev-Zel'dovich effect.  Collisional excitation cooling is likely
to be dominant in the others (circles and crosses). Ultra-violet photons
from these SNRs may contribute to reionizing the IGM. Free-free emission
is stronger in halos not blown away by SN explosion (crosses).  They may
be the first X-ray sources in the universe.

So far we have focused our attention on the evolution of SNRs until
substantial gas cooling takes place within them. It is well known that
expanding and cooling gas shells are subject to both thermal and
gravitational instabilities (Ostriker \& Cowie 1981; Vishniac 1983;
MacLow \& Norman 1993). Simple analytic arguments show that the
fragmentation of the swept-up shell occurs efficiently when the ambient
density is higher than $\sim 10$ cm$^{-3}$ for the explosion energy of
$10^{51}-10^{52}$ erg (Salvaterra, Ferrara \& Schneider 2004).
Following Machida et al. (2005), we have compared the local free-fall
time-scale
\begin{equation}
t_{\rm ff}=\sqrt{\frac{3\pi}{32 G\rho_{\rm sh}}},
\end{equation}
and the sound crossing time-scale 
\begin{equation}
t_{\rm sc}=\frac{\Delta r_{\rm sh}}{c_{\rm s}},
\end{equation}
where $\rho_{\rm sh}$ is the mean mass density, $\Delta r_{\rm sh}$ is
the thickness and $c_{s}$ is the sound speed, of the thin shell. We
define the shell in our simulations as the region within the full width
at half maximum (FWHM) of the density peak, and assume that it is stable
against fragmentation if $t_{\rm ff} > t_{\rm sc}$.

We find that the above stability condition is satisfied in all the
`blown-away' cases with the initial I-front expansion (triangles and
circles in Figure \ref{fig:em}). In these runs, the initial gas density
is lower than 1 cm$^{-3}$ due to radiative feedback, resulting in the
low efficiencies of both cooling and deceleration of the SN shocks. For
the run plotted in Figure \ref{fig:prof1} ($M_{\rm h}=3.2 \times 10^5
\msun$ and $E_{\rm SN}=10^{51}$ erg), the shell attains the highest
density at $t \sim 3 \times 10^5$ yr with the FWHM $\Delta r_{\rm sh}
\sim 0.5$ pc resolved with eight gas meshes. At this point, the above
time-scales are $t_{\rm ff} \sim 2 \times 10^7$ yr and $t_{\rm sc} \sim 6
\times 10^4$ yr, respectively.  On the other hand, the shell may become
unstable in some `stalled' cases, in which the shell expansion slows
down soon after the internal thermal energy is lost and cannot
effectively counteract contraction.  For the run plotted in Figure
\ref{fig:prof3} ($M_{\rm h}= 10^7\msun$ and $E_{\rm SN}=10^{53}$ erg),
$t_{\rm ff}$ becomes shorter than $t_{\rm sc}$ when an extremely dense
shell forms at $t=80$ yr.  Note that including molecular hydrogen
cooling would further promote fragmention by lowering the temperature
and increasing the density of the shell.  Precise account of this effect
needs further study and we address it in greater detail in a separate
paper (Kitayama \& Yoshida, in preparation).  Finally, it is worth
noting that first SN explosions could trigger self-propagating
star-formation (e.g. Dopita 1985; Efremov \& Elmegreen 1998).  In large
halos, second-generation stars could form in SN shells in the {\it
induced} manner, rather than via gas condensation onto the halo center.

\section{Discussion}

The first generation stars cause three key feedback effects on
subsequent star/galaxy formation: (1) Radiative feedback by
photo-ionizing, -heating, and -dissociating the IGM as well as
primordial gas clouds, (2) Mechanical feedback by expelling the
surroundings, and (3) Chemical feedback by distributing heavy elements.

Radiative feedback usually has negative effects on star-formation
efficiency. The gas in low-mass ($M_{\rm h} \simlt 10^6 \msun$) halos is
almost completely ionized even by a single massive star (Paper
I). Soft-UV photons in the Lyman-Werner bands can dissociate hydrogen
molecules, the main coolant in the primordial gas, in slightly larger
mass halos. While Paper I confirmed a positive feedback effect on H$_2$
formation just ahead of the I-front (Ricotti et al. 2001), it is shown
to be rather weak and tentative.

It has also been argued that soft X-rays increase the ionization
fraction and promote the formation of hydrogen molecules via the H$^{-}$
path (Haiman, Abel \& Rees 2001; Oh 2001; Venkatesan, Giroux \& Shull
2001). Heating by secondary electrons may, however, be substantial and
the net effect of soft X-rays is still unclear.  Indeed,
3-D adaptive-mesh-refinement simulations by Machacek et
al. (2003) showed that, whereas molecular hydrogen formation is promoted
in dense gas clouds {\it that have already formed}, the dominant effect
of the X-ray background is to heat the IGM.  Machacek
et al. also conclude that the net effect (either positive or negative)
is found to be quite small for X-ray intensities below $\sim 10^{-23}$
erg s$^{-1}$ cm$^{-2}$ Hz$^{-1}$ sr$^{-1}$.  From our estimates of the
X-ray emission efficiency of a single SNR, this level of flux is
achieved if more than a few thousand SN explosions occurred in a
comoving cubic mega-parsec volume by $z\sim 15$. For reasonable early
star-formation models, this rate is implausibly high, and thus we
conclude that global feedback effects by X-ray emission from early SNRs
are unimportant.

We further argue that the expected intensity of H$_2$ lines from an
early SNR is rather small. Consider as an optimistic limit that the
ejected energy of $E_{\rm SN}=10^{51}$erg is used only to heat the
ambient gas to $T=8000$K, so that the atomic cooling is negligible.  The
heated region should enclose the gas mass of $5 \times 10^5 \msun$ or the
total mass of $ 3 \times 10^6 \msun$.  It takes $\sim 10^7$ yr for the
shock with $T = 8000$K to propagate over the virial radius $\sim 220$ pc
of such a halo at $z\sim 20$. The expected optimal luminosity is then
$L_{\rm H2} \sim 3 \times 10^{36}$ erg s$^{-1}$.  In practice, some of
the gas is heated to higher temperatures, atomic cooling will dominate
over H$_2$ lines, and the luminosity should be lower than this value.

Mechanical feedback from the first stars is often cited as a destructive
process in the context of early structure formation. The energy released
by a single Population III SN explosion can be as large as $\sim
10^{53}$ erg, which is much larger than the gravitational binding energy
of a minihalo with mass $\sim 10^6 M_{\odot}$.  At the first sight, this
simple argument appears to support the notion that high-$z$ SNe are
enormously destructive.  As we have explicitly shown, however, the
evolution of the {\it cooling} remnants crucially depends on the
properties of the surrounding medium, particularly on the central
density after the gas is re-distributed by radiation.

The efficiency of radiative feedback declines sharply above a certain
halo mass $M_{\rm h} \sim 10^7 \msun$ (Paper I), so does that of the
mechanical feedback by SNe. Halos above this threshold can thus be ideal
candidates for the hosts of the first massive black-holes in cases where
supernovae leave remnant black-holes.  Our calculations indicate that the
mass deposition rate at parsec scales reaches $\sim 0.1\msun$yr$^{-1}$
in the case of $E_{\rm SN}=10^{53}$ erg, $M_{\rm h}=10^7\msun$ at
$t=10^7$ yr (Sec. \ref{sec:ev}).  This value is much higher than a
simple estimation for the Bondi accretion rate $4 \times 10^{-8} (M_{\rm
BH}/100 \msun)^2 \msun$ yr$^{-1}$ for the ambient gas density $10^4$
cm$^{-3}$ and temperature $10^4$K, or the rate resulting in the Eddinton
luminosity, $2 \times 10^{-6} (\epsilon/0.1)^{-1}(M_{\rm BH}/100 \msun)
\msun$ yr$^{-1}$, where $M_{\rm BH}$ is the black-hole mass and
$\epsilon$ is the efficiency factor of fueling. Therefore, it should at
least provide a necessary (obviously not sufficient) condition for the
near-Eddinton growth of black-holes at the center of these large halos
even after the mechanical feedback from the first SNe.  We note,
however, that actual mass accretion around black holes, occurring at
much smaller radius than our simulations probe, is much more complicated
because of radiation-hydrodynamical effects, non-spherical geometry,
etc.  These issues are still highly uncertain and beyond the scope of
the present paper.

As has been often discussed in the literature (Ricotti \& Ostriker 2004;
Yoshida et al. 2004), the first star formation is likely to be episodic,
at least locally, because the feedback effects tend to quench further
star-formation in the same place.  Although metal-enrichment by the
first stars could greatly enhance the gas cooling efficiency, which
would then change the mode of star-formation to that dominated by
low-mass stars (Mackey, Bromm \& Hernquist 2003; Bromm \& Loeb 2003),
the onset of this 'second-generation' stars may be delayed particularly
in low-mass halos. Our results imply that the strength of this negative
feedback substantially decreases for $\simgt 10^7 M_{\odot}$ halos.  In
CDM models, halos grow hierarchically and thus {\it all} the halos with
$\simgt 10^7 M_{\odot}$ used to be `mini-halo' at some earlier epoch. If
star-formation is prevented earlier in the progenitor small halos either
by rapid mass accretion (Yoshida et al. 2003) or by a soft-UV background
which raises the minimum mass scale for the gas cloud formation
(Machacek et al 2001; Wise \& Abel 2005), primordial stars could form in
$\simgt 10^7 M_{\odot}$ halos.  Only if multiple explosions take place
after a burst of star-formation in the large halos, a complete
destruction can take place.  Wada \& Venkatesan (2003) showed that such
multiple explosions can easily disrupt pre-galactic disks.

\section{Conclusions}

First supernova explosions in the early universe have a number of
consequences on subsequent star formation.  While they tend to expel gas
from low-mass halos, the extent to which such feedback can operate
depends sensitively on the initial configuration of the medium
surrounding the explosion site. We have shown explicitly that the
ionization front expansion around a massive progenitor star can
significantly aid gas evacuation by supernova blastwaves.  For halos
with mass $\simlt 10^6 \msun$, essentially all the gas is expelled by
merely a tenth of nominal SN explosion energy.  On the other hand, if
the \HII region is confined within the central portion of the halo, the
gas density remains high, and the necessary SN energy for `blow-out' is
larger by more than two orders of magnitude than the total binding
energy for baryons in the halo, i.e., $E_{\rm SN} > 10^{54}$ erg, for
$M_{\rm h} = 10^7 \msun$ at $z=20$.  This is due to very high efficiency
of radiative cooling in the dense gas clouds.

We have also found that there are mainly three radiative processes that
are important in the early SNRs.  (i) Collisional excitation cooling,
particularly via the Ly$\alpha$ line emission, is dominant for either
higher-mass halos or lower SN energy.  (ii) Compton cooling is dominant
for either lower-mass halos or higher SN energy.  (iii) Free-free
emission gets stronger for massive ($>10^7\msun$) halos, in which the
expansion of an \HII region is greatly suppressed, and can make them
compact X-ray sources.

Overall, the emergence of the first generation stars have significant
impacts on the thermal state of the IGM in the early universe in at
least two ways; the initially neutral cosmic gas is first photo-ionized
and heated to $\sim 10^4$ K by stellar radiation, and subsequently
heated to even higher temperature and enriched with heavy elements by
SNe.  Our results imply that radiative and SN feedback processes are
closely related to each other and thus need be treated in a
self-consistent manner. Because of the strong feedback effects, early
star-formation is likely to be self-regulating; if the first stars are
massive, only one period of star-formation is possible for a small halo
and its descendant.  The sharp decline in the efficiency of both
feedback effects at $M_{\rm h} >10^7 \msun$ indicates that the emergence
of large halos can drastically raise the global star formation rate of
the Universe. This may lead to accelerating the metal enrichment of the
IGM at an early stage of galaxy formation.

\hfill 

We thank Satoru Ikeuchi, Kuniaki Masai and Paul Shapiro for fruitful
discussions, and the referee for useful comments.  NY acknowledges
support from the 21$^{\rm st}$ COE Program ORIUM at Nagoya
University. This work is supported in part by the Grants-in-Aid by the
Ministry of Education, Science and Culture of Japan (14740133:TK).

\appendix

\section{Basic features of the SNR evolution in the early universe}

The evolution of a SNR may be divided into three stages: (1) a
free-expansion phase when the SN ejecta expand at nearly a constant
velocity, (2) an adiabatic Sedov phase which is described by a
self-similar solution, and (3) a pressure-driven snow-plow phase when a
dense shell forms within which the thermal energy of the gas is radiated
away. In the following, we describe basic elements of physical processes
associated with point explosions.

The free-expansion phase lasts until the ejecta sweeps up roughly
the same amount of mass as their own, $M_{\rm ej}$, in the surrounding medium: 
\begin{equation}
  M_{\rm ej} = \int_0^{R_{\rm sw}} dr ~ 4\pi r^2 m_p n_{\rm H}/X,   
\end{equation}
yielding 
\begin{equation}
R_{\rm sw} \sim \left(\frac{3M_{\rm ej}X}{4\pi m_p n_{\rm H}}\right)^{1/3}
= 11 {\rm pc} \left(\frac{n_{\rm H}}{{\rm cm}^{-3}}\right)^{-1/3}
\left(\frac{M_{\rm ej}}{200 M_{\odot}} \right)^{1/3},
\end{equation}
where $R_{\rm sw}$ is the radius enclosing $M_{\rm ej}$, $m_p$ is the
proton mass, $X$ is the hydrogen mass fraction, and $n_{\rm H}$ is the
mean hydrogen number density of the surrounding medium.  With the
ejected kinetic energy $E_{\rm SN}$, the initial shock velocity is
\begin{equation}
V_{\rm si} \sim  \sqrt{\frac{2 E_{\rm SN}}{M_{\rm ej}}} = 710 ~ 
{\rm km s}^{-1}  \left(\frac{M_{\rm ej}}{200 M_{\odot}} \right)^{-1/2}
\left(\frac{E_{\rm SN}}{10^{51}{\rm erg}} \right)^{1/2}, 
\label{eqvs}
\end{equation}
corresponding to the initial shock temperature of 
\begin{equation}
T_{\rm si} = \frac{3\mu m_p V_{\rm si}^2}{16 k} = 6.7 \times 10^6 
~{\rm K}  \left(\frac{M_{\rm ej}}{200 M_{\odot}} \right)^{-1}
\left(\frac{E_{\rm SN}}{10^{51}{\rm erg}} \right), 
\label{eqtsi}
\end{equation}
where $\mu$ is the mean molecular weight and $k$ is the Boltzmann
constant. The time taken for the shock to reach $R_{\rm sw}$ is then
estimated as
\begin{equation}
t_{\rm sw}\sim  \frac{R_{\rm sw}}{V_{\rm si}} =
1.6 \times 10^4 ~{\rm yr}
\left(\frac{n_{\rm H}}{{\rm cm}^{-3}}\right)^{-1/3}
\left(\frac{M_{\rm ej}}{200 M_{\odot}} \right)^{5/6}
\left(\frac{E_{\rm SN}}{10^{51}{\rm erg}} \right)^{-1/2}.
\label{eqsw}
\end{equation}
The time required for electrons to attain the shock temperature via
Coulomb collisions is roughly (Masai 1994)
\begin{equation}
t_{\rm s} \sim 6 \times 10^3 ~{\rm yr}
\left(\frac{n_{\rm H}}{{\rm cm}^{-3}}\right)^{-1}. 
\left(\frac{T_{\rm si}}{10^{7}{\rm K}} \right)^{3/2},
\label{eqts}
\end{equation}
So long as $t_{\rm s} < t_{\rm sw}$, one can assume that electron-ion
equilibration is achieved and $T_{\rm e} \sim T_{\rm si}$, where $T_{\rm
e}$ is the electron temperature. The hot gas in the interior of high
redshift ($z\ga 10$) SNRs cools initially via free-free emission and
Compton scattering with the time-scales given respectively by
\begin{eqnarray}
t_{\rm ff} &\sim& 10^7  ~{\rm yr} 
\left(\frac{n_{\rm H}}{{\rm cm}^{-3}}\right)^{-1} 
\left(\frac{T_{\rm e}}{10^{7}{\rm K}} \right)^{1/2}, 
\label{eqtc1}\\ 
t_{\rm Comp} &\sim& 7 \times 10^6 ~{\rm yr}
\left(\frac{1 + z}{20}\right)^{-4}. 
\label{eqtc2}
\end{eqnarray}
If either $t_{\rm ff}$ or $t_{\rm Comp}$ is shorter than $t_{\rm sw}$,
the ejected energy is radiated away before entering the Sedov phase. If
$t_{\rm ff}$ and $t_{\rm Comp}$ are both greater than $t_{\rm sw}$, the
evolution is adiabatic.  Note that $t_{\rm s}$ can be larger than
$t_{\rm sw}$ for low $n_{\rm H}$ or high $E_{\rm SN}$. The radiative
cooling, however, is unimportant in such cases and the shock can still
be well described by the adiabatic solution and electron-ion
equilibration is eventually achieved.

\epsscale{0.85}
\begin{figure}
\plotone{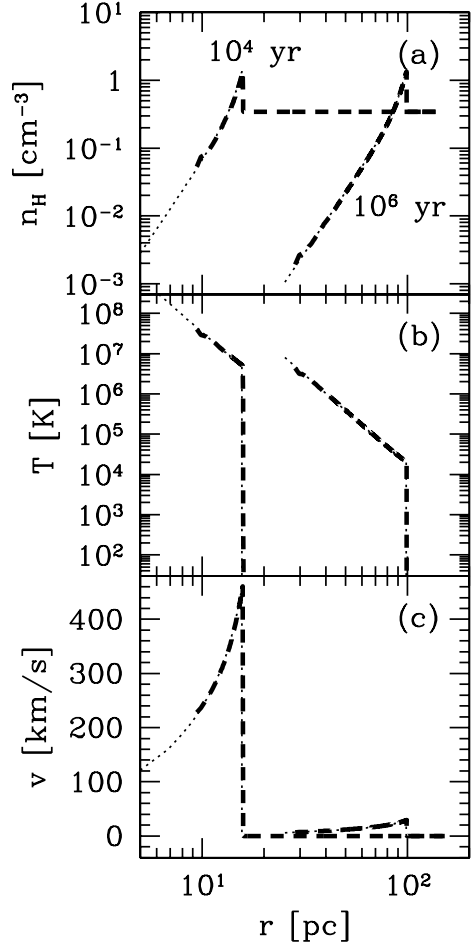} \caption{Propagation of the blastwave with $E_{\rm
SN}=10^{51}$ erg in a uniform medium corresponding to a virialized halo
with $M_{\rm h}=10^{6}$ M$_\odot$ and $z_c=20$. Dashed lines indicate
the simulation results at $10^4$ and $10^6$ yr, while dotted lines
denote the corresponding self-similar solutions.  \label{fig:sedov1}}
\end{figure}

For a power-law density profile with $n_{\rm H} \propto r^{-w}$, the
expansion of a blastwave in the adiabatic Sedov phase is described by
the self-similar solution,
\begin{equation}
R_{\rm s}(t) \propto  \left(\frac{E_{\rm SN}\;t^2}{n_{\rm H}}\right)^{1/(5-w)},
\label{eq:sedov_p}
\end{equation}
where $R_{\rm s}(t)$ is the shock radius at time $t$ after the explosion
(Sedov 1959; see also Ostriker \& McKee 1988). In reality, the gas
density profile can be more complex because photo-evaporation due to
radiation from the central star is very effective in low-mass halos
(Paper I). If the size of the \HII region is comparable to the virial
radius (e.g. Fig. \ref{fig:prof1}), the gas density inside the \HII
region is nearly uniform at $ n_{\rm H}\simeq 0.3 
[(1+z)/20]^3 ~{\rm cm}^{-3}$.  For a constant density
medium, the Sedov-Taylor self-similar solution reduces to
\begin{eqnarray}
R_{\rm s}(t) &\simeq& 32 ~{\rm pc} 
\left(\frac{n_{\rm H}}{{\rm cm}^{-3}}\right)^{-1/5}
\left(\frac{E_{\rm SN}}{10^{51}{\rm erg}} \right)^{1/5}
\left(\frac{t}{10^5{\rm yr}}\right)^{2/5}, 
\label{eq:sedovr}\\
T_{\rm s}(t) &\simeq& 2.1\times 10^5 ~{\rm K} 
\left(\frac{n_{\rm H}}{{\rm cm}^{-3}}\right)^{-2/5}
\left(\frac{E_{\rm SN}}{10^{51}{\rm erg}} \right)^{2/5}
\left(\frac{t}{10^5{\rm yr}}\right)^{-6/5},
\label{eq:sedovt}
\end{eqnarray}
Note that the shock radius in equation (\ref{eq:sedov_p}) scales only
weakly with the density and the explosion energy unless the density
profile is very steep.  The 3-D simulations of Bromm et
al. (2003) show that the evolution of the shock radius is well described
by equation (\ref{eq:sedovr}) in a piecewise manner for regions having
density profiles with different slopes.

The above adiabatic solution breaks down once the gas becomes
radiative. A dense shell will form roughly at the sag time (Cox 1972;
Shull 1980) at which $t \sim t_{\rm ff}/10$ with $T_e = T_{\rm s}(t)$
from equations (\ref{eqtc1}) and (\ref{eq:sedovt});
\begin{eqnarray}
t_{\rm sg} \sim 10^{5} ~ {\rm yr} 
\left(\frac{n_{\rm H}}{{\rm cm}^{-3}}\right)^{-3/4}
\left(\frac{E_{\rm SN}}{10^{51}{\rm erg}} \right)^{1/8}.
\label{eq:tsg}
\end{eqnarray}
The shock will then follow the snow-plow solution, $R_{\rm s}(t) \propto
t^{2/7}$ (Shull 1980).  The momentum-driven slow expansion is still
important over a cosmological time-scale (Bertschinger 1983), with the
radius doubling every twenty folds in time.

\begin{figure}
\plotone{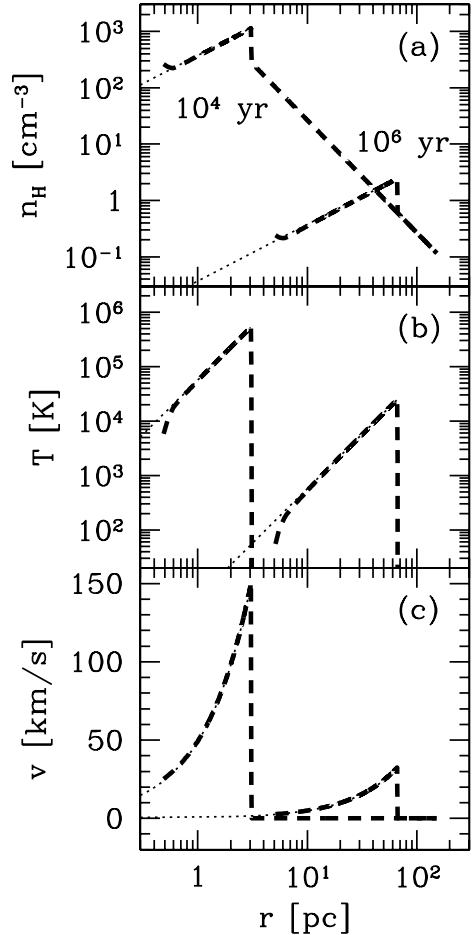} \caption{Same as Figure \ref{fig:sedov1} except that
the medium has a gas density gradient with index $w=2$.
\label{fig:sedov2}}
\end{figure}

\section{Code test}

We have tested and verified the validity of our Lagrangian code by a
variety of test problems. The results of the `Str\"omgren sphere' test
is presented in Paper I. We here describe the results of the Sedov
explosion problem (Sedov 1959), which is the most relevant to the
present paper.

We solve the evolution of a blastwave in a spherical halo with $M_{\rm
h}=10^6 \msun$ and $z_{\rm c}=20$.  For the sake of direct comparisons
with the self-similar solutions, we neglect gravitational force and
radiative cooling. The initial gas density profile is a power-law
$n_{\rm H} \propto r^{-w}$ with either $w=0$ (uniform) or $w=2$.  The
total number of gas shells is $N=1000$; the inner-most shell encloses
the gas mass $1 \msun$ and the shell mass increases by a constant ratio,
$1\%$, between the adjacent shells toward the envelope.  The thermal
energy of $E_{\rm SN}=10^{51}$ erg is given to the inner-most shell.
The initial thermal energy in the other shells is essentially zero. The
initial velocity is also taken to be zero. The simulation is otherwise
performed in the same way as the other runs in the present paper.

Figures \ref{fig:sedov1} and \ref{fig:sedov2} compare the outputs of the
above test runs and analytical solutions (Ostriker \& McKee 1988).  The
numerical results reproduce well the latter for both $w=0$ and
$w=2$. The positions of shock front, defined as the peak of thermal
pressure, agree with the analytical expectations within 1\%.  This
further ensures the capability of the code for the simulations in the
present paper.


\begin{thebibliography}{11}

\bibitem{ABN} 
Abel, T., Bryan, G. L., \& Norman, M. L. 2002, 
Science, 295, 93

\bibitem{pisn}
Barkat, Z., Rakavy, G., \& Sack, N.
1967, Phys. Rev. Lett., 18, 379

\bibitem{berc}
Bertschinger, E.
1983, ApJ, 268, 17

\bibitem{berch}
Bertschinger, E. 1985, ApJS, 58, 39

\bibitem{bac}
Bond, J. R., Arnett, W. D., \& Carr, B. J.
1984, ApJ, 280, 825

\bibitem{BW} 
Bowers, R. L. \& Wilson, J. R.,
1991, 
Numerical Modelling in Applied Physics
and Astrophysics (Boston: Jones and Bartlett)

\bibitem{Br}
Bromm, V., Coppi, P. S., \& Larson, R. B.
2002, ApJ, 564, 23

\bibitem{BL03}
Bromm, V., \&  Loeb, A.  2003, Nature, 425, 812

\bibitem{BSN}
Bromm, V., Yoshida, N., \& Hernquist, L.
2003,
ApJ, 596, L135 

\bibitem{bullock} 
Bullock, J., et al. 
2001, 
MNRAS, 321, 559

\bibitem{art}
Caramana, E.J., Shashkov, M.J., \& Whalen, P. P.,
1998,
Journal of Computational Physics, 144, 70

\bibitem{carr}
Carr, B. J., Bond, J. R., \& Arnett, W. D.,  1984, 
ApJ, 277, 445


\bibitem{Couchman}
Couchman, H. M. P., \& Rees, M. J.
1986, MNRAS, 221, 53

\bibitem{Cox72}
Cox, D. P. 1972, ApJ, 178, 159 

\bibitem{ds86}
Dekel, A., \& Silk, J.
1986, ApJ, 303, 39

\bibitem{dopita}
Dopita, M. A. 1985, ApJ, 295, L5

\bibitem{ee}
Efremov, Y. N., \& Elmegreen, B. G.
1998, MNRAS, 299, 643

\bibitem{FWH}
Fryer, C. L., Woosley, S. E., \& Heger, A.
2001, ApJ, 550, 372

\bibitem{FurL03}
Fukugita, M. \& Kawasaki, M.
1994, MNRAS, 343, L25

\bibitem{haiman3} 
Haiman, Z., Abel, T., \& Rees, M. J., 
2000, 
ApJ, 534, 11

\bibitem{HW}
Heger, A., \& Woosley, S. E.
2002, ApJ, 567, 532


\bibitem{i81}
Ikeuchi, S.,
1981, Pub. Astr. Soc. Japan, 33, 211

\bibitem{KI00} 
Kitayama, T. \& Ikeuchi, S.,
2000,
ApJ, 529, 615

\bibitem{KSUI00} 
Kitayama, T., Tajiri, Y., Susa, H., Umemura, M., \& Ikeuchi, S.,
2000,
MNRAS, 315, L1

\bibitem{KSUI} 
Kitayama, T., Susa, H., Umemura, M., \& Ikeuchi, S.,
2001,
MNRAS, 326, 1353

\bibitem{Ki04} 
Kitayama, T., Yoshida, N., Susa, H., \& Umemura, M., 
2004, ApJ, 613, 631 (Paper I)

\bibitem{MBA}
Machacek, M. E., Bryan, G. L., \& Abel, T.
2001, ApJ, 548, 509

\bibitem{MBAII}
Machacek, M. E., Bryan, G. L., \& Abel, T.
2003, MNRAS, 338, 273

\bibitem{HD}
Machida, M. N., Tomisaka, K., Nakamura, F. \& Fujimoto, M.
2005, ApJ,  622, 39

\bibitem{Mac}
Mac Low, M.-M., \& Norman, M. L.
1993, ApJ, 407, 207

\bibitem{MB}
Mackey, J., Bromm, V., \& Hernquist, L.
2003, ApJ, 586, 1

\bibitem{MFR}
Madau, P., Ferrara, A., \& Rees, M. J.
2001, ApJ, 555, 92

\bibitem{masai}
Masai, K. 1994, ApJ, 437, 770

\bibitem{mori}
Mori, M., Ferrara, A.,  \& Madau, P., 2002,  ApJ, 571, 40 

\bibitem{nfw} 
Navarro, J. F., Frenk, C. S. \& White, S. D. M.,
1997,
ApJ, 490, 493

\bibitem{oh}
Oh, S. P. 2001, ApJ, 569, 558

\bibitem{ock}
Oh, S. P., Cooray, A., \& Kamionkowski, M.
2003, MNRAS, 342, 20

\bibitem{OP}
Omukai, K., \& Palla, F.
2003, ApJ, 589, 677

\bibitem{OC}
Ostriker, J. P., \&  Cowie, L. L.  1981, ApJ, 243, L127

\bibitem{OM}
Ostriker, J. P., \& McKee, C. F.
1988, Rev. Mod. Phys., 60, 1

\bibitem{qian}
Qian, Y.-Z. \& Wasserburg, G. J.,
2005, ApJ in press

\bibitem{ricottia}
Ricotti, M., Gnedin, N. Y., \& Shull, J. M.
2001, ApJ, 560, 580

\bibitem{ricottib}
Ricotti, M., \& Ostriker, J. P.
2004, MNRAS, 350, 539

\bibitem{salvaterra}
Salvaterra, R., Ferrara, A., \&  Schneider, R.  2004, New Astronomy, 10, 113

\bibitem{sch02}
Schaerer, D. 2002, A\&A, 382, 28

\bibitem{raf}
Schneider, R., Ferrara, A., Ntarayan, P. \& Omukai, K. 2002,
ApJ, 571, 30

\bibitem{sedov}
Sedov, L. I.
1959, Similarity and Dimensional Methods in Mechanics (New York: Academic Press)

\bibitem{shull}
Shull, J. M.  1980, ApJ, 237, 769

\bibitem{shullmackee}
Shull, J. M. \& McKee, C. F. 1979, ApJS, 227, 131

\bibitem{tegmark93}
Tegmark, M.,  Silk, J., \&  Evrard, A.   1993, ApJ, 417, 54 

\bibitem{tegmark}
Tegmark, M., Silk, J., Rees, M. J., Blanchard, A., Abel, T.,
\& Palla, F. 1997, ApJ, 474, 1

\bibitem{ter}
Terlevich, R., Tenorio-Tagle, G., Franco, J., \& Melnick, J.
1992, MNRAS, 255, 713

\bibitem{umeda}
Umeda, H., \& Nomoto, K.
2002, ApJ, 565, 385

\bibitem{umeda03}
Umeda, H., \& Nomoto, K.
2003, Nature, 422, 871

\bibitem{VEN} 
Venkatesan, A., Giroux, M. \& Shull, J. M., 
2001, ApJ, 563, 1

\bibitem{vishniac}
Vishniac, E. T.  1983,  ApJ, 274, 152

\bibitem{vishniac85}
Vishniac, E. T., Ostriker, J. P., \& Bertschinger, E.,
1985, ApJ, 291, 399

\bibitem{wada}
Wada, K., \& Venkatesan, A.
2003, ApJ, 591, 38

\bibitem{whalen}
Whalen, D., Abel, T., \& Norman, M. N.
2004, ApJ, 610, 14

\bibitem{wan}
Wandel, A.
1985, ApJ, 294, 385

\bibitem{wise}
Wise, J. \& Abel, T.
2005, astro-ph/0411558

\bibitem{paperI} 
Yoshida, N., Abel, T., Hernquist, L. \& Sugiyama, N., 
2003, ApJ, 592, 645

\bibitem{ymetal} 
Yoshida, N., Bromm, V., \& Hernquist, L. 
2004, ApJ, 605, 579

\end{thebibliography}
\end{document}